\documentclass[9pt,twocolumn,twoside]{osajnl}

\journal{ol} 

\setboolean{shortarticle}{true}

\usepackage{siunitx}

\usepackage{lineno}
\usepackage{pdfpages}
\usepackage{xcolor}
\usepackage{dblfloatfix}    

\title{Multi-gigawatt peak power post-compression in a bulk multi-pass cell at high repetition rate}

\author[1]{Ann-Kathrin Raab}
\author[2]{Marcus Seidel}
\author[1]{Chen Guo}
\author[1]{Ivan Sytcevich}
\author[3]{Gunnar Arisholm}
\author[1]{Anne L'Huillier}
\author[1]{Cord L. Arnold}
\author[1,2,*]{Anne-Lise Viotti}

\affil[1]{Department of Physics, Lund University, P.O. Box 118, SE-22100 Lund, Sweden}
\affil[2]{Deutsches Elektronen-Synchrotron DESY, Notkestraße 85, 22607 Hamburg, Germany}
\affil[3]{FFI (Norwegian Defence Research Establishment), P.O. Box 25, NO-2027, Kjeller, Norway}

\affil[*]{Corresponding author: anne-lise.viotti@fysik.lth.se}

\begin{abstract}
The output of a $\SI{200}{\kilo\hertz}$, $\SI{34}{\watt}$, $\SI{300}{\femto\second}$ Yb amplifier is compressed to $\SI{31}{\femto\second}$ with $>\SI{88}{\percent}$ efficiency to reach a peak power of $\SI{2.5}{\giga\watt}$, which to date is a record for a single-stage bulk multi-pass cell. Despite operation 80 times above the critical power for self-focusing in bulk material, the setup demonstrates excellent preservation of the input beam quality. Extensive beam and pulse characterizations are performed to show that the compressed pulses are promising drivers for high harmonic generation and nonlinear optics in gases or solids.

\end{abstract}

\setboolean{displaycopyright}{true}

\begin{document}
\maketitle

High peak power ultrafast sources are key enabling tools for time-resolved studies at the femto- and atto-second time scales. In addition to increased repetition rates, long-term stability and  excellent beam quality are highly desirable, especially for time-resolved pump-probe experiments or nanoscale imaging \cite{herbert2021probing}.
Few-fs pulses with multi-$\SI{}{\micro\joule}$ level energies
are needed to drive high harmonic generation (HHG) for applications demanding high repetition rate, e.g.~coincidence spectroscopy or photoelectron microscopy \cite{mikaelsson2021high, klas2021ultra}. 
To generate such optical pulses, Ytterbium (Yb) based sources are gaining popularity as they show excellent power and repetition rate scalability \cite{russbueldt2014innoslab}. They also benefit from simpler thermal management due to the small quantum defect and efficient, cheap diode pumping schemes thanks to the long upper-state lifetime. However, their gain bandwidth
does not allow pulse durations smaller than few hundreds of fs. Shorter pulse durations can be accessed via optical parametric chirped pulse amplification \cite{dubietis1992powerful}, though at the cost of conversion efficiency and complexity. 

A viable, much more efficient alternative is direct pulse compression by spectral broadening via self-phase modulation (SPM) in a Kerr medium followed by chirp removal \cite{nagy2021high}.
One recent approach, based on multi-pass cell (MPC) geometry, displays exceptional average power handling capabilities for post-compression \cite{hanna2021nonlinear,viotti2022multi}. 
MPCs have been routinely operated with $\SI{100}{\watt}$ to $\SI{}{\kilo\watt}$ power levels \cite{grebing2020kilowatt}. Moreover, MPCs have been used with pulse energies ranging from a few $\SI{}{\micro\joule}$ to more than $\SI{100}{\milli\joule}$, resulting in peak powers approaching the $\SI{}{\tera\watt}$ regime \cite{kaumanns2021spectral}. For $\SI{}{\giga\watt}$ input peak powers, gas-filled MPCs have been mainly utilized: they readily enable nonlinearity tuning and typically exhibit > $\SI{90}{\percent}$ power transmission. However, they require bulky, costly vacuum equipment and cannot be operated above the critical power of the nonlinear gas medium, which is why
using a bulk material in an MPC is a much simpler alternative.

Bulk MPCs have been operated with $\SI{}{\mega\watt}$ input peak powers and, at best, post-compression up to $\sim\SI{1}{\giga\watt}$ peak power was reached \cite{omar2021gw,seidel2022factor}. Interestingly, the input peak powers surpass the critical power of the bulk nonlinear material multiple times. 
For example, a previous bulk MPC study at $\SI{1.03}{\micro\meter}$ used $\SI{150}{\mega\watt}$ input peak power and reported a compression ratio of $11$ \cite{omar2021gw}, while Vicentini et al.~reached  a compression factor of $3$ in a second broadening stage with $\SI{240}{\mega\watt}$ \cite{vicentini2020nonlinear}. 
These correspond to input peak powers exceeding the critical power of fused silica ($\approx\SI{4.3}{\mega\watt}$, \cite{fibich2000critical}) by a factor 30 and 50, respectively. A larger input (output) peak power, $\SI{280}{\mega\watt}$ ($\SI{440}{\mega\watt}$), was obtained in a factor 2.8 self-compression experiment \cite{jargot2018self} performed at $\SI{1.55}{\micro\meter}$. This setting exceeded the critical power 30 times and translates to $\SI{200}{\mega\watt}$ output peak power at $\SI{1.03}{\micro\meter}$, according to the wavelength squared scaling \cite{fibich2000critical}. 
Operating in such supercritical self-focusing regime bears the risk that the spatial nonlinearities cause pulse quality degradation, resulting in limited compressibility and the emergence of strong pulse pedestals. Moreover, for peak powers greater than $\SI{50}{}$ times the critical power, small scale self-focusing and multi-filament breakup can occur \cite{campillo2009small}.
It has recently been shown that using multiple, thin Kerr media instead of one single, thicker medium allows cleaner compression and scaling of compression ratios by distributing the Kerr nonlinearity along the MPC \cite{seidel2021ultrafast}. When using >$\SI{100}{\mega\watt}$ input peak power, it is, however, questionable if multiple thin plates suffice to suppress spatio-temporal couplings. 

In this work, we present an extensive characterization of a bulk MPC with, to the best of our knowledge, the highest input and output peak powers, operating >80 times above the critical power of fused silica. SPM in thin plates allows us to reach $\SI{31}{\femto\second}$ pulses with $\SI{2.5}{\giga\watt}$ at $\SI{200}{\kilo\hertz}$ and a compression factor of $\approx\SI{10}{}$. This setup, solely built from readily available components, stands out due to its simplicity, overall cost, power efficiency and small footprint. The influence of the input pulse dispersion and temporal structure on the spectral broadening in the MPC is investigated. Spectral, power and carrier-envelope phase (CEP) stability measurements are carried out.
In spite of the high input peak power, the measured spatial, spectral and temporal quality of the post-compressed pulses is excellent.
This is essential when employing these pulses as drivers for secondary sources, e.g.~HHG or other frequency conversion processes.

\begin{figure}[t]
\centering
\includegraphics[width=0.8\linewidth]{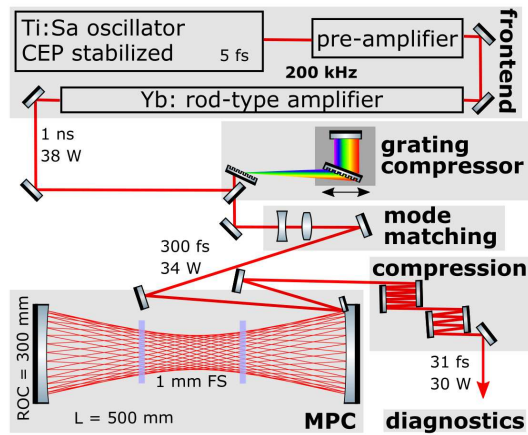}
\caption{Schematic of the laser architecture and the experimental setup with: motorized grating compressor, mode-matching, MPC and compression.}
\label{fig:setup}
\end{figure}

\begin{figure}[t]
\centering
\includegraphics[width=0.95\linewidth]{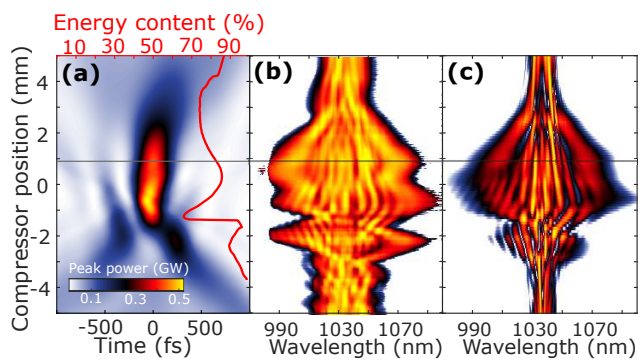}
\caption{(a) Reconstructed temporal profile of the MPC input pulse via d-scan with the relative energy content in the main pulse (red). The grey horizontal line represents the input compression position used in the study. (b) Measured and (c) simulated broadened MPC spectrum as a function of the relative grating compressor position. (b) and (c) are normalized.}
\label{fig:rod}
\end{figure}

The experimental setup is depicted in Fig.~\ref{fig:setup}. The frontend is a CEP-stabilized Titanium:Sapphire (Ti:Sa) oscillator. A narrow part of its spectrum is temporally stretched with a chirped fiber Bragg grating and amplified at $\SI{200}{\kilo\hertz}$ in an Yb rod-type amplifier. The output is compressed by a transmission grating compressor to $\SI{300}{\femto\second}$ full-width-half-maximum (FWHM) with $\SI{170}{\micro\joule}$. In the compressor, the second grating and the retro-reflector are mounted on a motorized stage, which allows fine-tuning of the spectral phase by varying the group delay and third order dispersion at a rate of -$\SI{29000}{\femto\second\squared\per\milli\meter}$ and $\SI{160000}{\femto\second\cubed\per\milli\meter}$, respectively. In the following, the grating compressor output is referred to as the input to the MPC. A lens telescope 
matches the beam to the eigenmode of an Herriott-type MPC. The cell is designed for 15 roundtrips using standard $\SI{1030}{\nano\meter}$ quarter-wave stack mirrors with a radius of curvature (ROC) of $\SI{300}{\milli\meter}$. The cell length is $\SI{500}{\milli\meter}$ and the Kerr media are two $\SI{1}{\milli\meter}$ thin anti-reflection (AR) coated fused silica plates spaced by $\SI{15}{\centi\meter}$ and symmetrically placed in the MPC (see Fig.~\ref{fig:setup}). In- and out-coupling is done via a scraper mirror. The SPM-induced chirp is removed via chirped mirrors, which compensate for a total dispersion of $\SI{2800}{\femto\second\squared}$. Adjusting the positions of the plates, and therefore the peak intensities in the bulk media, allows us to tune the broadening in the cell. This gives flexibility to operate the MPC at a targeted spectral broadening for a range of input peak powers. 

The MPC input pulses are characterized by the dispersion scan (d-scan) technique \cite{sytcevich2021characterizing}. The dispersion is varied by moving the motorized grating, as indicated by the double-headed arrow in Fig.~\ref{fig:setup}. The retrieved d-scan trace (see Supplement 1) allows the reconstruction of the input pulse at different compressor positions, as shown in Fig.~\ref{fig:rod} (a). In Fig.~\ref{fig:rod}, the relative position from the center (zero) of the scanning range of the compressor stage is the common y-axis. Positive (negative) compressor positions correspond to a positive (negative) net input pulse dispersion, respectively. The relative main pulse energy contained in twice the FWHM, and integrated over the full measurement window of \textpm $\SI{2.5}{\pico\second}$, is compared to the total pulse energy (red curve in Fig.~\ref{fig:rod} (a)). Larger values of the energy content correspond to cleaner input pulses with minimized pedestals, while shorter input pulses with slightly higher peak powers exhibit strong double-pulse structures and lower energy content.
While changing the dispersion, the spectrum of the compressed MPC output is measured (see Fig.~\ref{fig:rod} (b)). This study is interesting as the input pulse parameters for optimum spectral broadening and clean compression are not usually obvious and do not necessarily correspond to the shortest input pulse. Three cases can then be identified: Around +$\SI{0.9}{\milli\meter}$, the input pulse has reduced pedestals, a large energy content in the main pulse ($\SI{82}{\percent}$), while having a rather short duration ($\SI{300}{\femto\second}$) and a high peak power ($\SI{370}{\mega\watt}$). At this position, highlighted by the horizontal line across Fig.~\ref{fig:rod}, the pulse is slightly positively chirped and the resulting SPM spectrum is broad. At -$\SI{0.7}{\milli\meter}$, the spectral broadening is the largest as the input pulse is the shortest, with a higher peak power. However, the side pulse becomes strong enough to broaden as well, thus leading to spectral fringes observed in Fig.~\ref{fig:rod} (b). 
Finally, substantial broadening is also noticed around -$\SI{2}{\milli\meter}$. There, the input pulse is a negatively chirped double pulse for which both pulses spectrally broaden independently, leading to strong modulations in the spectrum of Fig.~\ref{fig:rod} (b).

Additionally, the experimental data is supported by simulations, using the SISYFOS (SImulation SYstem For Optical Science) code \cite{arisholm2021simulation} for nonlinear beam propagation in the MPC pictured in Fig.~\ref{fig:setup}. The reflectivity and dispersion of the MPC mirrors, the transmission of the AR-coating of the thin plates are included in the simulations, as well as the Kerr nonlinearity of air, owing to the large peak intensities in the center of the cell. While the input spatial beam is a fundamental Gaussian with $M^2=1$, the retrieved spectrum and phase from the d-scan measurements are used as input to the simulations. Figure~\ref{fig:rod} (c) shows the simulated spectral broadening scan. The main spectral features from the experiment (Fig.~\ref{fig:rod} (b)) are well reproduced in the simulations and the positions of the optimum compression and the largest broadening points match well. For negatively chirped input pulses (from -$\SI{0.7}{\milli\meter}$ to around -$\SI{2.5}{\milli\meter}$), the spectrum is strongly modulated, leading to complex post-compressed simulated pulse profiles (see Supplement 1), unsuitable for the envisioned ultrafast applications. The combination of the motorized grating compressor together with a spectrometer recording the output of the MPC constitutes a helpful tool to determine the optimum input pulse dispersion regime, in our case slightly positive, at +$\SI{0.9}{\milli\meter}$.

\begin{figure}[t]
\centering
\includegraphics[width=0.85\linewidth]{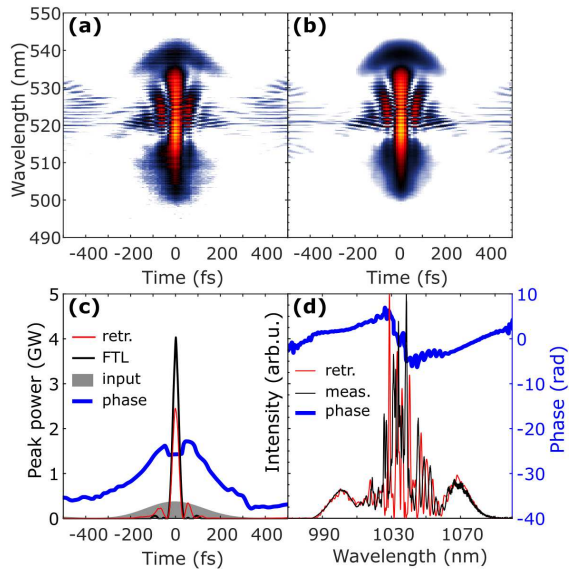}
\caption{(a) Measured and (b) retrieved FROG traces in logarithmic scale. (c) Temporal profile of the compressed MPC output (red, $\SI{31}{\femto\second}$), compared to the FTL (black, $\SI{30}{\femto\second}$) and the MPC input (grey, $\SI{300}{\femto\second}$). (d) Retrieved and measured spectra. The fast spectral modulations and the significant peak power difference with respect to the transform limited pulse originate from a pre-pulse $\SI{1}{\pico\second}$ away (see Supplement 1).}
\label{fig:frog}
\end{figure}

The MPC input, as determined in Fig.~\ref{fig:rod}, has an average power of $\SI{34}{\watt}$ and a peak power of $\SI{370}{\mega\watt}$ with a duration of $\SI{300}{\femto\second}$ (FWHM). The post-compressed output power is $\SI{30}{\watt}$, which translates to an overall transmission efficiency above $\SI{88}{\percent}$. The compressed MPC output pulses are characterized by a second harmonic frequency-resolved optical gating (FROG) setup, for which the measured and retrieved traces are shown in Fig.~\ref{fig:frog} (a) and (b).  The retrieved FWHM of $\SI{31}{\femto\second}$, normalized via the measured output pulse energy of $\SI{150}{\micro\joule}$, yields a peak power of $\SI{2.5}{\giga\watt}$ (see Fig.~\ref{fig:frog} (c)), so far the highest reported value for a bulk MPC. Measured and retrieved broadened spectra match well, as shown in Fig.~\ref{fig:frog} (d), and correspond to a Fourier transform limit (FTL) of $\SI{30}{\femto\second}$ FWHM, which is the bandwidth limit supported by the current MPC mirrors. The compression factor is $9.7$. The corresponding simulated broadened spectrum, extracted from Fig.~\ref{fig:rod} (c) at +$\SI{0.9}{\milli\meter}$, matches the results of the FROG measurement very well (see Supplement 1). Similar to Fig.~\ref{fig:rod} (a), the efficiency of compression is assessed by estimating the relative energy content in the main peak of the post-compressed output pulse, selecting twice the FWHM FTL over an integration window of \textpm $\SI{2.5}{\pico\second}$. $\SI{77}{\percent}$ of the energy remains in the main pulse, similar to recent results in gas-filled MPCs \cite{viotti2021temporal}.


During time-resolved pump-probe experiments, the long-term stability of the compressed MPC output matters. In terms of average power, a $\SI{1}{\hour}$ measurement with a $\SI{10}{\hertz}$ sampling rate shows that the MPC does not increase the fluctuations significantly, from $\SI{0.27}{\percent}$ input root-mean-square error (RMSE) to $\SI{0.32}{\percent}$ output. A parallel measurement of the broadened spectrum over $\SI{1}{\hour}$, see Supplement 1, reveals a standard deviation of the FTL of $\SI{0.2}{\femto\second}$. 

\begin{figure}[t]
\centering
\includegraphics[width=0.95\linewidth]{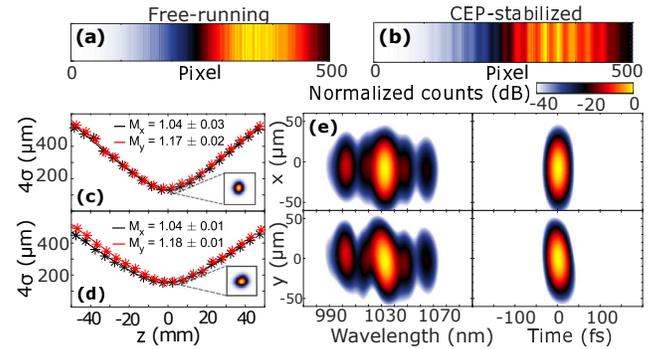}
\caption{$\SI{1}{\second}$ time-averaged spectral fringes (linear scale) from an f-2f interferometer, comparing the oscillator in (a) a free-running state and (b) actively CEP-stabilized. $M^2$ measurements (c) before the mode-matching unit and (d) for the compressed MPC output, with insets of the normalized beam profiles in the focus. (e) Reconstructed spectral and temporal distributions in $x$ and $y$ directions.
}
\label{fig:quality}
\end{figure}

For phase-sensitive experiments such as HHG, CEP stability has a strong influence. While not essential for the current pulse duration of $\SI{31}{\femto\second}$, it becomes important for further compression. In our case, only the oscillator is CEP-stabilized. To measure the CEP stability, an f-2f interferometer is used at the output of the MPC, where the second harmonic of a white light generated in a sapphire crystal interferes with the blue edge of the same white light. The resulting spectral fringes are recorded in a single-shot measurement with a line camera \cite{mikaelsson2021high}, where the read-out speed allows us to capture each pulse at $\SI{200}{\kilo\hertz}$. While the phase is rather noisy, the time-average spectra over $\SI{1}{\second}$, shown in Fig.~\ref{fig:quality} (a) and (b), are clearly different when the oscillator is free-running and when it is actively stabilized. Further numerical investigations with Allan variance analysis validate this distinction (see Supplement 1). Observing fringes in the stabilized case means that the oscillator's phase is preserved through the entire amplifier chain, which includes a pre-amplifier and a rod-type amplifier with a large stretching/compression ratio (see Fig.~\ref{fig:setup}), as well as the MPC. A correction loop can eventually be implemented to stabilize the phase drift \cite{natile2019cep}. 



Finally, a beam quality assessment is performed, starting with an $M^2$ measurement to judge the focusability, which is highly relevant for, e.g., HHG. Figures~\ref{fig:quality} (c) and (d) compare the $M^2$ at the input and output of the MPC with no significant change: a value below $1.2$ is obtained for both spatial directions.
Since high peak intensities are reached in the cell, concerns arise regarding eventual spatio-temporal couplings. In fact, the MPC input peak power of $\SI{370}{\mega\watt}$ exceeds the critical power of fused silica >80 times, largely above any previously reported result in bulk MPCs \cite{omar2021gw,vicentini2020nonlinear}.


Full spatio-spectral/temporal 3D characterization was previously performed for gas-filled MPCs operating close to the critical power and yielded Strehl ratios around 0.9 \cite{daher2020multipass}. In this work, a full 3D characterization employing spatially-resolved Fourier transform spectrometry \cite{miranda2014spatiotemporal} is conducted. 
This method allows us to spectrally resolve the wavefront and the beam profile of the output of the MPC. Together with the reference pulse measurement obtained by FROG, the pulse is numerically focused and reconstructed in both spectral and temporal domains, as shown in Fig.~\ref{fig:quality} (e). In the x-direction, the spectrum is homogeneous but a slight spatial chirped is observed in the y-direction. This translates into a pulse front-tilt in the time domain for the same direction. An identical measurement performed at the MPC input gives similar results and also displays spatial chirp (see Supplement 1), indicating that the MPC does not introduce significant spatio-temporal couplings. The origin of such chirp is most likely a slight misalignment of the retro-reflector in the grating compressor. From the results of this study, the 3D time profiles can be compared to an ideal wavefront-compensated pulse leading to a $\SI{20}{\percent}$ difference. This can be partly explained by the pulse front-tilt in the $y$ direction.
Accounting for the FROG retrieval, the total 3D Strehl ratio is: $S_{3D}=0.69$. Disregarding the temporal domain allows us to extract the commonly used 2D Strehl ratio, which is $S_{2D}=0.89$. 

In conclusion, we present, to the best of our knowledge, a bulk MPC at $\SI{200}{\kilo\hertz}$ with the highest peak powers so far achieved: $\SI{370}{\mega\watt}$ input with $\SI{300}{\femto\second}$ compressed down to $\SI{31}{\femto\second}$ with $\SI{2.5}{\giga\watt}$ output. This single-stage compression setup is power-, space- and cost-efficient, being solely realized with off-the-shelf optics. The operating pulse parameter regime demonstrated here directly competes with standard gas-filled hollow-core fibers used for spectral broadening \cite{kottig2017generation}. Despite a working point above 80 times the critical self-focusing power for fused silica, no spatio-temporal couplings are introduced, as the full 3D characterization shows.  This MPC is also a suitable first stage towards the few-cycle regime and can be inserted into a cascaded spectral broadening scheme with, e.g., a second cell \cite{muller2021multipass} or a capillary \cite{lavenu2019high}. Additionally, increasing pulse energies can enable further peak power scaling. As mentioned previously, the position of the Kerr medium can be tuned and the setup size can be geometrically scaled up, although only until a certain practical limit \cite{viotti2022multi}. Moreover, for high input peak powers, another limit will be the critical power in air, making it difficult to operate without a chamber. Methods to circumvent size and peak power restrictions have been recently demonstrated in gas-filled MPCs, such as utilization of higher-order spatial modes \cite{kaumanns2021spectral}, multiplexing \cite{stark2022divided} or bow-tie type cavities \cite{heyl2022high}. Together with preserved beam quality, power, spectral and phase stability, this setup constitutes a promising route for driving applications demanding large peak powers and high repetition rates. The source has been recently used to generate high-order harmonics in argon and neon up to cut-off energies equal to $\SI{80}{\electronvolt}$ and $\SI{135}{\electronvolt}$, respectively.
 
\begin{backmatter}
\bmsection{Funding} We acknowledge support from the European Research Council (Advanced grant QPAP, 884900); the Swedish Research Council (2019-06275, 2016-04907, 2013-8185); the Wallenberg Center for Quantum Technology funded by the Knut and Alice Wallenberg Foundation; the Crafoord Foundation (20200584) and Lund Laser Centre.

\bmsection{Disclosures} The authors declare no conflicts of interest.

\bmsection{Data availability} Data underlying the results presented in this paper are not publicly available at this time but may be obtained from the authors upon reasonable request.

\bmsection{Supplemental document}
See Supplement 1 for supporting content. 

\end{backmatter}


\begin{thebibliography}{10}
	\newcommand{\enquote}[1]{``#1''}
	
	\bibitem{herbert2021probing}
	G.~Herbert, A.~Wöste, J.~Vogelsang, T.~Quenzel, D.~Wang, P.~Groß, and
	C.~Lienau, {\protect\JournalTitle{ACS Photonics}} \textbf{8}, 2573 (2021).
	
	\bibitem{mikaelsson2021high}
	S.~Mikaelsson, J.~Vogelsang, C.~Guo, I.~Sytcevich, A.-L. Viotti, F.~Langer,
	Y.-C. Cheng, S.~Nandi, W.~Jin, A.~Olofsson, R.~Weissenbilder, J.~Mauritsson,
	A.~L'Huillier, M.~Gisselbrecht, and C.~L. Arnold,
	{\protect\JournalTitle{Nanophotonics}} \textbf{10}, 117 (2021).
	
	\bibitem{klas2021ultra}
	R.~Klas, A.~Kirsche, M.~Gebhardt, J.~Buldt, H.~Stark, S.~H{\"a}drich,
	J.~Rothhardt, and J.~Limpert, {\protect\JournalTitle{PhotoniX}} \textbf{2}, 1
	(2021).
	
	\bibitem{russbueldt2014innoslab}
	P.~Russbueldt, D.~Hoffmann, M.~H{\"o}fer, J.~L{\"o}hring, J.~Luttmann,
	A.~Meissner, J.~Weitenberg, M.~Traub, T.~Sartorius, D.~Esser, R.~Wester,
	P.~Loosen, and R.~Poprawe, {\protect\JournalTitle{IEEE Journal of Selected
			Topics in Quantum Electronics}} \textbf{21}, 447 (2014).
	
	\bibitem{dubietis1992powerful}
	A.~Dubietis, G.~Jonusauskas, and A.~Piskarskas, {\protect\JournalTitle{Optics
			Communication}} \textbf{88}, 437 (1992).
	
	\bibitem{nagy2021high}
	T.~Nagy, P.~Simon, and L.~Veisz, {\protect\JournalTitle{Advances in Physics:
			X}} \textbf{6}, 1845795 (2021).
	
	\bibitem{hanna2021nonlinear}
	M.~Hanna, F.~Guichard, N.~Daher, Q.~Bournet, X.~Délen, and P.~Georges,
	{\protect\JournalTitle{Laser Photonics and Reviews}} \textbf{15}, 2100220
	(2021).
	
	\bibitem{viotti2022multi}
	A.-L. Viotti, M.~Seidel, E.~Escoto, S.~Rajhans, W.~P. Leemans, I.~Hartl, and
	C.~M. Heyl, {\protect\JournalTitle{Optica}} \textbf{9}, 197 (2022).
	
	\bibitem{grebing2020kilowatt}
	C.~Grebing, M.~Müller, J.~Buldt, H.~Stark, and J.~Limpert,
	{\protect\JournalTitle{Optics Letters}} \textbf{45}, 6250 (2020).
	
	\bibitem{kaumanns2021spectral}
	M.~Kaumanns, D.~Kormin, T.~Nubbemeyer, V.~Pervak, and S.~Karsch,
	{\protect\JournalTitle{Optics Letters}} \textbf{46}, 929 (2021).
	
	\bibitem{omar2021gw}
	A.~Omar, S.~Ahmed, M.~Hoffmann, and C.~J. Saraceno, \enquote{{GW} peak power,
		sub-30-fs pulses from efficient single-stage pulse compressor at 400-{kHz},}
	in \emph{Conference on Lasers and Electro-Optics,}  (Optica Publishing Group,
	2021), p. STh2I.4.
	
	\bibitem{seidel2022factor}
	M.~Seidel, P.~Balla, C.~Li, G.~Arisholm, L.~Winkelmann, I.~Hartl, and C.~M.
	Heyl, {\protect\JournalTitle{Ultrafast Science}} \textbf{2022}, 9754919
	(2022).
	
	\bibitem{vicentini2020nonlinear}
	E.~Vicentini, Y.~Wang, D.~Gatti, A.~Gambetta, P.~Laporta, G.~Galzerano,
	K.~Curtis, K.~McEwan, C.~R. Howle, and N.~Coluccelli,
	{\protect\JournalTitle{Optics Express}} \textbf{28}, 4541 (2020).
	
	\bibitem{fibich2000critical}
	G.~Fibich and A.~L. Gaeta, {\protect\JournalTitle{Optics Letters}} \textbf{25},
	335 (2000).
	
	\bibitem{jargot2018self}
	G.~Jargot, N.~Daher, L.~Lavenu, X.~Delen, N.~Forget, M.~Hanna, and P.~Georges,
	{\protect\JournalTitle{Optics Letters}} \textbf{43}, 5643 (2018).
	
	\bibitem{campillo2009small}
	R.~Boyd, S.~Lukishova, and Y.~Shen, eds., \emph{{Self-focusing: Past and
			Present}}, vol. 114 of \emph{Topics in {Applied} {Physics}} (Springer, New
	York, NY, 2009).
	
	\bibitem{seidel2021ultrafast}
	M.~Seidel, F.~Pressacco, O.~Akcaalan, T.~Binhammer, J.~Darvill, N.~Ekanayake,
	M.~Frede, U.~Grosse-Wortmann, M.~Heber, C.~M. Heyl, D.~Kutnyakhov, C.~Li,
	C.~Mohr, J.~Müller, O.~Puncken, H.~Redlin, N.~Schirmel, S.~Schulz,
	A.~Swiderski, H.~Tavakol, H.~Tünnermann, C.~Vidoli, L.~Wenthaus, N.~Wind,
	L.~Winkelmann, B.~Manschwetus, and I.~Hartl, {\protect\JournalTitle{Laser \&
			Photonics Reviews}} \textbf{16}, 2100268 (2021).
	
	\bibitem{sytcevich2021characterizing}
	I.~Sytcevich, C.~Guo, S.~Mikaelsson, J.~Vogelsang, A.-L. Viotti, B.~Alonso,
	R.~Romero, P.~T. Guerreiro, {\'I}.~J. Sola, A.~L’Huillier, H.~Crespo,
	M.~Miranda, and C.~L. Arnold, {\protect\JournalTitle{Journal of the Optical
			Society of America B}} \textbf{38}, 1546 (2021).
	
	\bibitem{arisholm2021simulation}
	G.~Arisholm and H.~Fonnum, \enquote{Simulation system for optical science
		({SISYFOS}) - tutorial, version 2,} Tech. rep., ser. FFI-rapport. Norwegian
	Defense Research Establishment (FFI) (2021).
	
	\bibitem{viotti2021temporal}
	A.-L. Viotti, S.~Alisauskas, H.~Tünnermann, E.~Escoto, M.~Seidel, K.~Dudde,
	B.~Manschwetus, I.~Hartl, and C.~M. Heyl, {\protect\JournalTitle{Optics
			Letters}} \textbf{46}, 4686 (2021).
	
	\bibitem{natile2019cep}
	M.~Natile, A.~Golinelli, L.~Lavenu, F.~Guichard, M.~Hanna, Y.~Zaouter,
	R.~Chiche, X.~Chen, J.~F. Hergott, W.~Boutu, H.~Merdji, and P.~Georges,
	{\protect\JournalTitle{Optics Letters}} \textbf{44}, 3909 (2019).
	
	\bibitem{daher2020multipass}
	N.~Daher, F.~Guichard, S.~W. Jolly, X.~Delen, F.~Quere, M.~Hanna, and
	P.~Georges, {\protect\JournalTitle{JOSA B}} \textbf{37}, 993 (2020).
	
	\bibitem{miranda2014spatiotemporal}
	M.~Miranda, M.~Kotur, P.~Rudawski, C.~Guo, A.~Harth, A.~L’Huillier, and C.~L.
	Arnold, {\protect\JournalTitle{Optics Letters}} \textbf{39}, 5142 (2014).
	
	\bibitem{kottig2017generation}
	F.~Köttig, F.~Tani, C.~M. Biersach, J.~C. Travers, and P.~S.~J. Russell,
	{\protect\JournalTitle{Optica}} \textbf{4}, 1272 (2017).
	
	\bibitem{muller2021multipass}
	M.~Müller, J.~Buldt, H.~Stark, C.~Grebing, and J.~Limpert,
	{\protect\JournalTitle{Optics Letters}} \textbf{44}, 2678 (2021).
	
	\bibitem{lavenu2019high}
	L.~Lavenu, M.~Natile, F.~Guichard, X.~Delen, M.~Hanna, Y.~Zaouter, and
	P.~Georges, {\protect\JournalTitle{Optics Express}} \textbf{27}, 1958 (2019).
	
	\bibitem{stark2022divided}
	H.~Stark, C.~Grebing, J.~Buldt, M.~Müller, A.~Klenke, and J.~Limpert,
	\enquote{Divided-pulse nonlinear compression in a multipass cell,} in
	\emph{Fiber Lasers XIX: Technology and Systems,} , vol. 11981 (SPIE, 2022),
	p. 1198101.
	
	\bibitem{heyl2022high}
	C.~M. Heyl, M.~Seidel, E.~Escoto, A.~Schönberg, S.~Carlström, G.~Arisholm,
	T.~Lang, and I.~Hartl, {\protect\JournalTitle{Journal of Physics: Photonics}}
	\textbf{4}, 014002 (2022).
	
\end{thebibliography}
\end{document}